\documentclass[apj]{emulateapj}

\usepackage{xspace,graphics,graphicx}
\usepackage{natbib}
\newcommand{\mkms}{{\rm \; km\;s^{-1}}}

\newcommand{\nlris}{101 }

\slugcomment{Submitted to ApJL} 

\shorttitle{Cool, Metal-Enriched Accretion at $z \sim 0.5$}
\shortauthors{Rubin et al.}

\begin{document}

\title{The Direct Detection of Cool, Metal-Enriched Gas Accretion onto Galaxies at $z \sim 0.5$}
\author{Kate H. R. Rubin \altaffilmark{1}, 
J. Xavier Prochaska \altaffilmark{2}, 
David C. Koo \altaffilmark{2} \& 
Andrew C. Phillips \altaffilmark{2}}
\altaffiltext{1}{Max-Planck-Institut f\"ur Astronomie, K\"onigstuhl 17, 69117 Heidelberg, Germany; rubin@mpia.de}
\altaffiltext{2}{Department of Astronomy \& Astrophysics, UCO/Lick Observatory, University of California, 1156 High St, Santa Cruz, CA 95064}

\begin{abstract}
We report on the discovery of cool gas inflow toward six star-forming galaxies with redshifts $z\sim0.35 - 1$.  Analysis of \ion{Mg}{2} and \ion{Fe}{2} resonance-line absorption in Keck/LRIS spectroscopy of this sample reveals velocity shifts of $80 - 200 \mkms$ and equivalent widths for inflowing gas of $\gtrsim 0.6$ \AA \ in five of the six objects.  The host galaxies exhibit a wide range in star formation rates (SFR $\sim 1 - 40~M_{\odot}~\mathrm{yr}^{-1}$) and have stellar masses similar to that of the Milky Way ($\log M_*/M_{\odot} \sim 9.6 - 10.5$).  Imaging from the \emph{Hubble Space Telescope} Advanced Camera for Surveys indicates that five of the six galaxies have highly inclined ($i > 55^{\circ}$), disk-like morphologies.  These data represent the first unambiguous detection of inflow into isolated, star-forming galaxies in the distant universe.  We suggest that the inflow is due to the infall of enriched material from dwarf satellites and/or a galactic fountain within the galaxies.  
Assuming that the material has been enriched to $0.1Z_{\odot}$ 
and has a physical extent approximately equal to that of the galaxies, we infer mass inflow rates of $dM_\mathrm{in}/dt \gtrsim 0.2 - 3~M_{\odot}~\mathrm{yr}^{-1}$ for four of these systems.   
Finally, from comparison of these absorption lines to the profiles of \ion{Mg}{2} and \ion{Fe}{2} absorption in a larger spectroscopic sample of $\sim 100$ objects, we measure a covering fraction of cool inflow of at least $6\%$, but cannot rule out the presence of enriched infall onto as many as $\sim 40$ of these galaxies. 

\end{abstract}

\section{Introduction}\label{sec.intro}

The accretion of gas onto galaxies is regarded as a process fundamental to their formation \citep{Larson1972,ReesOstriker1977} and is required to reconcile the limited cool gas supply in galactic disks with the cosmic star formation history \citep{Kennicutt1983,
Wong2004,Prochaska2005}.
Likewise, the star formation activity ($\sim1~M_{\odot}~\mathrm{yr}^{-1}$; \citealt{Robitaille2010}) in the inner Milky Way will consume the available gas on timescales of $\lesssim2$ Gyr, and can be maintained only if the local gas reservoir is replenished \citep{Blitz1996}.  
 Recent semi-analytic and cosmological hydrodynamic simulations \citep{Benson2003,Keres2005,Keres2009,DekelBirnboim2006} suggest that the 
 requisite baryons are delivered from the intergalactic medium (IGM) or low-mass satellites to high-redshift ($z \gtrsim 2$) galaxies via cool (temperature $\sim 10^4$ K) filaments of dense gas.  At $z < 1$, these filaments are truncated in massive halos, but accretion in the form of dense, cold clouds persists.
 
In the Milky Way, the accretion of cool gaseous material in high velocity clouds (HVCs) such as the Magellanic Stream is observed directly in 21 cm emission at distances of $5-20$ kpc  (\citealt{Wakker2001,Lockman2008,McClure2008}; Lehner \& Howk 2011).  
Previously-expelled material may be recycled to provide additional fuel for star formation (i.e., in a galactic fountain; \citealt{ShapiroField1976,Bregman1980,Marasco2011}), and likely gives rise to intermediate velocity clouds (IVCs) at distances $<5$ kpc \citep{Wakker2001,Bregman2009}. 
In addition, nearby spirals exhibit both extraplanar \ion{H}{1} clouds and morphological disturbances which may be attributed to gas infall \citep{Sancisi2008}. 
However, the emission from these diffuse structures is difficult to map beyond $\sim250$ Mpc \citep{ALFALFA2010}, 
and empirical evidence for gas accretion (or re-accretion) onto more distant galaxies is poignantly lacking.  

Cool accreting gas which has been enriched to modest metallicities (i.e., $Z\gtrsim0.1Z_{\odot}$) may give rise to absorption in background light sources in rest-frame ultraviolet transitions such as \ion{Mg}{2} $\lambda \lambda 2796, 2803$ or \ion{C}{4} $\lambda \lambda 1548, 1550$.  
However, studies of cool gas absorption along the sightlines toward star-forming galaxies have instead reported the ubiquity of outflows at $z\gtrsim1$ \citep{Weiner2009,RubinTKRS2009,Steidel2010}.
In one of the only studies showing evidence for inflows in distant systems, \citet{Sato2009} report \emph{redshifted} \ion{Na}{1} $\lambda \lambda 5890, 5896$ absorption in a sample of red, early-type objects, some of which exhibit line-emission from AGN activity.  Due to the low ionization potential of the \ion{Na}{1} ion (5.1 eV), however, HVC analogs rarely exhibit column densities $N_\mathrm{NaI}>10^{12.5}~\mathrm{cm}^{-2}$ \citep{BenBekhti2008,Richter2011}.
The moderate resolution and S/N spectra of \citet{Sato2009} are therefore likely to be sensitive to $N_\mathrm{HI}\gtrsim10^{20}~\mathrm{cm}^{-2}$ clouds, which are typically found  near (within $< 5$ kpc of) the Galactic disk \citep{Wakker2001,Richter2011}. 
\citet{LeFloch2007}, \citet{Ribaudo2011} and \citet{Giavalisco2011} have also recently reported evidence for cool accretion from absorption line analysis, though the latter two detections are tentative.

The lack of evidence for the inflow phenomenon is hardly surprising given that the predicted 
covering factor of accreting gas is small \citep[e.g., $< 10\%$ at $z \sim 1.5$;][]{Fumagalli2011}.  Further, as we will demonstrate, studies of cool gas kinematics in galaxy spectra are likely to identify inflows only if they achieve signal-to-noise (S/N) levels adequate for analysis of individual spectra, rather than coadded data. 
In this Letter, we report on high-S/N Keck/LRIS spectra of a sample of six star-forming galaxies at $0.35\lesssim z \lesssim1$ found to exhibit inflows traced by \ion{Mg}{2} and/or \ion{Fe}{2} $\lambda \lambda 2586, 2600$ absorption.  
We adopt a $\Lambda$CDM cosmology with $H_0=70~\rm km~s^{-1}~Mpc^{-1}$, $\rm \Omega_{M}=0.3$, and $\rm \Omega_{\Lambda} = 0.7$.  Magnitudes are given in the AB system.

\section{Observations}\label{sec.observations}

Our galaxy sample is drawn from a larger, magnitude-limited ($B_\mathrm{AB} < 23$) Keck/LRIS survey of cool gas kinematics in $\nlris$ galaxies at redshifts $0.3 < z < 1.4$ (Rubin et al.\ 2011, in prep) located in
fields imaged by the \emph{Hubble Space Telescope} Advanced Camera for Surveys
\citep[HST/ACS;][]{Giavalisco2004,Davis2007}.  
We derive rest-frame magnitudes and colors from these data and complementary ground-based optical\footnote{www.cfht.hawaii.edu/Science/CFHTLS-DATA/} and near-IR photometry  
\citep[Table~1;][]{Kajisawa2011,Wuyts2008} using the code KCORRECT \citep{Blanton2007}.

We obtained spectroscopy of this sample using the Low Resolution Imaging Spectrometer (LRIS) on Keck 1 \citep{Cohen1994}.    
We used $0.9\arcsec$ slitlets, and collected between four and eight $\sim 1800\rm~ sec$ exposures with FWHM $\sim 0.6 - 1.4 \arcsec$ seeing between 2008 May 30 UT and 2009 April 3 UT.  Our configuration of the two cameras with the 600/4000 grism, the 600/7500 grating, and the D560 dichroic provided FWHM~$\approx 200-400~ \rm km~s^{-1}$  and wavelength coverage $\lambda \sim 3200 - 8000$~\AA.  
The data were reduced using the XIDL LowRedux\footnote{http://www.ucolick.org/$\sim$xavier/LowRedux/} data reduction pipeline.

An iron-clad detection of inflow toward (or outflow from) a galaxy hinges on a precise and accurate determination of the systemic velocity.  We derived redshifts for the galaxies 
by calculating the best-fit lag between observed spectra and a linear combination of SDSS galaxy eigenspectra.  
We prefer redshift measurements based on stellar absorption, 
as they better trace the systemic velocity of the associated ensemble of dark matter and stars.
Therefore, where the stellar continuum S/N is adequate, we mask nebular emission lines in the data prior to redshift fitting.  
For EGS12027936, we adopt the redshift measured by the DEEP2 survey \citep{Davis2003}.
From a detailed 
analysis of our full LRIS sample, we find an RMS redshift uncertainty of $28\mkms$.  

\begin{figure}
\begin{center}
\includegraphics[angle=90,width=\columnwidth]{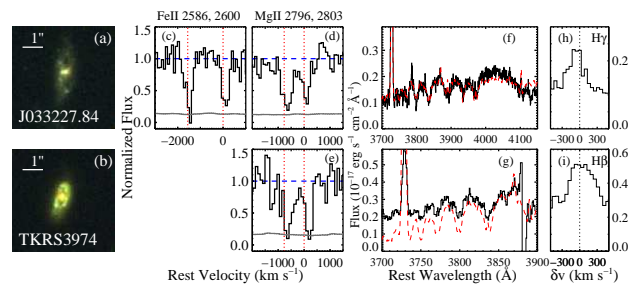}
\caption[]{Two galaxies with inflows measured with high confidence.  \emph{Panels (a) and (b):} $5\arcsec \times 5\arcsec$ ($\sim30\times30$ kpc) $BVi$ color HST/ACS images.  \emph{Panels (c), (d) and (e):} \ion{Fe}{2} and \ion{Mg}{2} transitions in the galaxy spectra.  Velocities are measured relative to the systemic velocities of the 2600 \AA \ line and the 2803 \AA \ line, respectively, as marked with vertical dotted lines.  Horizontal dashed lines mark the continuum level.  The gray lines show the $1\sigma$ error in each pixel.  \emph{Panels (f) and (g):} sections of the galaxy spectra showing higher-order Balmer transitions and [\ion{O}{2}] emission, with the fitted eigenspectrum template overlaid (red dashed line).  
\emph{Panels (h) and (i):} H$\gamma$ and H$\beta$ emission lines in the galaxy spectra.  This emission is offset from the systemic velocity by $\sim50\mkms$ in both galaxies. 
\label{fig.zproof}}
\end{center}
\end{figure}

\begin{figure}
\begin{center}
\includegraphics[angle=90,width=\columnwidth]{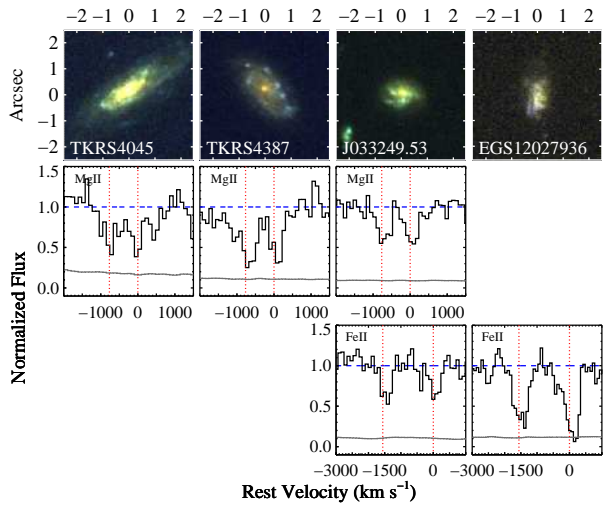}
\caption[]{The remaining four objects in our inflow sample.  \emph{Top:} $BVi$ color HST/ACS images.  \emph{Middle:} \ion{Mg}{2} transitions, with velocities measured relative to the 2803 \AA \ line.  \emph{Bottom:} \ion{Fe}{2} transitions, with velocities measured relative to the 2600 \AA \ line.  Colored curves are marked as in Figure~\ref{fig.zproof}.
\label{fig.inflows}}
\end{center}
\end{figure}

Figure~\ref{fig.zproof} demonstrates the results of the eigenspectra fits for two galaxies in our inflow sample.   
Note the offset of the nebular emission ($\sim 50\mkms$), which 
is indicative of a difference in the velocities and/or spatial distributions of stars and the interstellar medium (ISM) of these objects.  
Figure~\ref{fig.inflows} presents the images and absorption spectra for the 
remaining ``inflow'' galaxies.

\section{Absorption-Line Analysis}\label{sec.specanalysis}

To assess the signatures of cool gas outflow and inflow, we analyze
the line profiles of the \ion{Mg}{2} and \ion{Fe}{2}
transitions in our spectra (Rubin et al.\ 2011, in prep).  We construct two distinct models:
(1) a single-component model which assumes a Gaussian profile
(parameterized by a centroid, column density $N$, and $b$-value) with
a covering fraction $C_f$ independent of velocity; and (2) a two-component model with one component fixed at systemic and
having $C_f =1$. 

We assume that the likelihood function is given by the $\chi^2$ distribution for the model, and sample the posterior probability density function (PPDF) using the Multiple-Try Metropolis Markov Chain Monte Carlo technique \citep{Liu2000} as implemented in ROOT/RooFit, an object-oriented framework written in C++ \citep{BrunRademakers1997}.  Our code calculates the marginalized PPDF for each parameter and the equivalent width (EW) of the model absorption lines.  

This analysis has been implemented for our full spectroscopic sample.  While the majority of galaxies exhibit significantly blueshifted absorption indicative of outflows, the model fits for six galaxies indicate redshifted absorption with high probability.  That is, 
$>95\%$ of the marginalized PPDF for the one-component model
lies at $>0\mkms$ ($P_\mathrm{in,1}>0.95$) for both \ion{Mg}{2} and
\ion{Fe}{2} profiles where coverage is available.  
Table~1 reports $P_\mathrm{in}$ and the fitted velocity offsets
($\Delta v$) and EW for the one-component and
two-component models, subscripted with ``1" and ``2", respectively.

\section{Analysis of the Galaxies}\label{sec.galanalysis}

\begin{figure}
\begin{center}
\includegraphics[angle=90,width=\columnwidth]{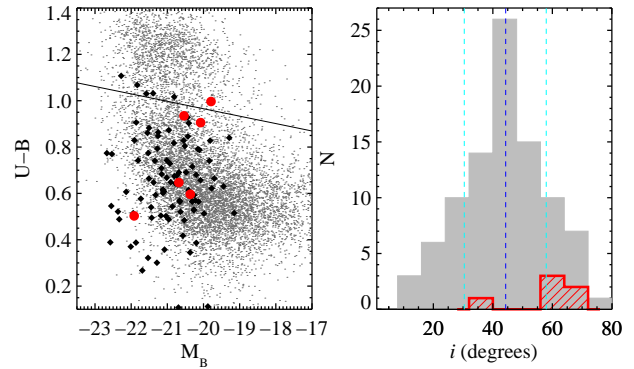}
\caption[]{\emph{Left:} Rest-frame color-magnitude diagram of our
  sample with inflows (red circles) and without (black diamonds). 
The solid line divides the
red sequence in the upper portion of the plot from the blue cloud below 
\citep{Willmer2006}.  
  Half of the inflow sample are star-forming galaxies in the blue cloud,
  while the other half lie in the green valley.  AEGIS galaxies
  with $0.3 < z < 1.05$ are also shown \citep[gray;][]{Davis2007}.
  \emph{Right:} The
  distribution of inclinations measured from HST/ACS
  imaging
  for the Rubin et al.\ (2011, in prep) sample (gray).  The median and
  $\pm 1\sigma$ 
  inclinations are marked with dashed lines.  
  The distribution of inclinations for
  galaxies exhibiting inflows (red) is skewed to high values (i.e., they are edge-on). 
\label{fig.cmdi}}
\end{center}
\end{figure}

Figure~\ref{fig.cmdi} shows the rest-frame colors and magnitudes of our six inflow galaxies (large red circles) and the parent spectroscopic sample 
(black diamonds; Rubin et al.\ 2011, in prep).  The inflow and parent samples occupy similar areas of the diagram.    
Three inflow objects lie in the blue cloud, and have star formation
rates (SFRs) $\sim10 - 40~M_{\odot}~\mathrm{yr}^{-1}$ and stellar
masses $\log M_*/M_{\odot} \sim9.6-10.1$ (Table 1).  The remaining
(redder) objects have low SFRs ($1-2~M_{\odot}~\mathrm{yr}^{-1}$) and lie in
the green valley between the red sequence and the blue cloud. 
Because these latter galaxies have disk-like rather than disturbed or early-type morphologies it is likely that these objects appear in the green valley due to enhanced dust reddening and their relatively modest SFRs, rather than the sudden cessation of star formation.  
Further, the HST/ACS imaging of these objects suggests the presence of
several compact, star-forming knots in the outskirts of the galactic
disks.   
Taken together, the stellar masses ($\log M_*/M_{\odot}\sim10.3 -10.5$), SFRs, optical colors and morphologies of these three galaxies
indicate that they are close analogs to the Milky Way, while the higher-SFR subset of the inflow sample 
has slightly lower stellar masses 
and more diffuse morphologies.

The distribution of inclinations measured from standard SExtractor
analysis \citep{BertinArnouts1996} among the parent spectroscopic sample (gray) is compared with that of the inflow sample (red) on the right side of Figure~\ref{fig.cmdi}.  All but one of the six inflow galaxies are highly inclined (having $i\gtrsim60^{\circ}$).   The Kolmogorov-Smirnov (K-S) test indicates only a 1.1\% probability that the inflow and parent populations are drawn from the same distribution.  
Five out of six inflow objects have inclinations close to (within $2^{\circ}$) or greater than the 84th-percentile value of inclination in the larger sample.  

\section{Discussion}\label{sec.discussion}

We have discovered a small sample of late-type, highly inclined, star-forming galaxies having SFRs over a wide range whose spectra show evidence for redshifted cool gas absorption.   
We interpret these kinematics as evidence for gaseous infall and  
discuss these findings in the context of galaxy formation models below, but
first critique alternative explanations.
If a galaxy's motions are dominated by rotation, the ISM absorption
may appear offset from the systemic velocity for an asymmetric gas distribution. 
Because this inflow galaxy sample is nearly edge-on, we are particularly
sensitive to velocity offsets in the direction of rotation.  \citet{Weiner2006} analyzed line emission in two-dimensional spectra of two of these objects, finding line-of-sight terminal rotation velocities of $64\mkms$ and $176\mkms$ for TKRS4045 and TKRS4387, respectively.  In an extreme scenario in which all \ion{Mg}{2}-absorbing ISM is located on only the receding side of these galaxies (and therefore has $C_f\sim0.5$), it may appear to be offset in velocity by $\sim30-90\mkms$ from the stellar absorption.  While most of our measured inflow velocities are much larger than $90\mkms$, TKRS4387 has a  
$\Delta v_2$ of only $\sim125\mkms$. 
However, given the velocity resolution and S/N of our spectra, we would be unlikely to detect this absorption unless it covers well over half ($\gtrsim70\%$) of the galaxy continuum (i.e., unless the profile is sufficiently deep).  
A spurious inflow signature could additionally result from 
an error in our determination of the galaxy systemic velocities due to, e.g.,  
a spatial offset between our slitlets and the centroid of the stellar continuum emission, such that the galaxy spectra are dominated by light from the approaching part of the disks.  
However, most of the galaxies have diameters $\lesssim1.6\arcsec$, and a $0.9\arcsec$ slit width was used in $\sim 1\arcsec$ seeing conditions; in addition, this scenario would yield an artificial inflow velocity of only $\sim30$ and $90\mkms$ in the cases of TKRS4045 and TKRS4387.  
We therefore adopt the interpretation that the absorption arises from metal-enriched gas flowing toward each galaxy from the IGM, as part of an accreting satellite, or from recycled wind material circulating in a galactic fountain.

\begin{figure}
\begin{center}
\includegraphics[angle=90,width=2in]{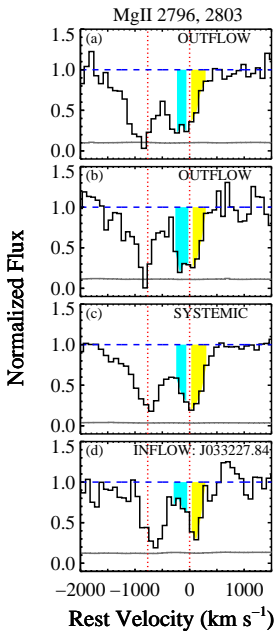}
\caption[]{
  Comparison of the \ion{Mg}{2} line profiles for a series of
  galaxies, each showing significant EWs ($>1.3$\AA) at $v>30 \mkms$ (yellow),
  with a wide range in blueshifted (cyan) and systemic absorption strengths.  
  Dashed, dotted, and gray curves are as in Figure 1.
  Despite the
  significant redshifted absorption, we only confidently identify
  inflow for the system without strong blueshifted/systemic absorption.  
\label{fig.ews}}
\end{center}
\end{figure}

At the most conservative level, the detection of six galaxies with inflows in a sample of \nlris (with sufficient S/N) implies a covering fraction of such material of $\approx 6\pm2\%$.  This estimate should be considered a firm lower limit, however.   
These six galaxies are unique not for the presence of strong absorption redward of systemic velocity, but instead for the \emph{absence} of strong, blueshifted absorption.  Figure~\ref{fig.ews} shows \ion{Mg}{2} line profiles for one object in our inflow sample (bottom panel) and three other objects drawn from the parent LRIS sample.  A by-eye analysis (and our fitting results) suggests that profiles (a) and (b) exhibit outflows; i.e., they show an excess of absorption blueward of systemic velocity.  Profile (c) is symmetric and dominated by absorption at systemic velocity.  However, all of these profiles have similar \ion{Mg}{2} $\lambda 2803$ EWs ($>1.3$ \AA) redward of systemic velocity (yellow), and thus could easily be tracing significant amounts of gas moving toward the host galaxies at $> 100\mkms$.  The difference in the measured kinematics is due to the differences in EWs at systemic velocity and blueward (cyan); i.e., profiles (a-c) have blueward EWs at least 0.5 \AA \ larger than the profile from the inflow sample (d).  We are therefore sensitive to inflows only in the absence of strong outflows, and are likely missing instances of cool accretion in our larger parent sample.  

For instance, the fraction of the parent sample with \ion{Fe}{2} $\lambda 2600$ or \ion{Mg}{2} $\lambda 2803$ EWs at least as large as those measured for our inflow sample in the velocity range $30\mkms<v<300\mkms$ (0.61 and 0.74 \AA, respectively) is 26 - 47\%.  Furthermore, 23\% of the parent sample has \ion{Mg}{2} $\lambda 2803$ EWs $> 0.37$ \AA \ in the velocity range $150\mkms < v < 300\mkms$ (i.e., at least as large as the EWs for those inflow galaxies with $P_\mathrm{in,2}>0.95$).  We therefore suspect the presence of inflow traced by saturated metal-line absorption in up to $\gtrsim 40\%$ of our parent sample, and suggest that it is likely occurring in at least $20\%$ of the galaxies.  
Finally, we note that due to the low frequency of detected inflows in individual spectra, the observed inflow signatures would likely be completely obscured in coadds of the parent sample (or subsamples thereof),  
similar to the composite spectra analyzed in \citet{Steidel2010}. 

Our observations provide almost no constraint on the distance between this inflowing gas and the galaxies or on the spatial distribution of the gaseous material.  However, 
the large model $C_f$ values ($\gtrsim0.7$) suggest that 
the material extends at minimum to sizes of order the size of the stellar disks (listed in Table 1).  In cases with high values of $P_\mathrm{in,2}$, we use the derived velocities and column densities to estimate a rough mass inflow rate.  We assume that the absorbing gas has a surface area given by $\pi R^2$, where $R$ is the average of the galaxy semi-major and semi-minor axes, and that the gas will accrete onto the galaxy with a timescale $\Delta v_2/R$.  
Neglecting ionization corrections, dust depletion, and assuming metallicities $Z = 0.1Z_{\odot}$, we find mass inflow rates $dM_\mathrm{in}/dt \sim0.2 - 3~M_{\odot}~\mathrm{yr}^{-1}$ (Table 1).  The values are slightly lower than the SFRs of the low-SFR half of the sample, and about an order of magnitude lower than the SFRs in the remaining galaxies.  
The inflow rates are also consistent with mass inflow rates derived for the Milky Way ($0.2~M_{\odot}~\mathrm{yr}^{-1}$; \citealt{Bregman2009}, Lehner \& Howk 2011, submitted).  

Given the 
strength of the metal-line absorption, the observed inflows are unlikely to arise from the ``cold flows" which are invoked to provide pristine hydrogen to star-forming galaxies from the IGM \citep{Fumagalli2011}.  Instead, this gas may have been enriched by star formation in satellite dwarf galaxies, or may have already cycled through the host galaxy's ISM.  
IVCs and HVCs in our own Galaxy could easily give rise to the inflow signatures observed in our sample, as they have a wide range of velocities (up to $> 300\mkms$) and are mostly optically thick in \ion{H}{1} with metallicities $\gtrsim0.1Z_{\odot}$ \citep{Wakker2001}.
The cosmological hydrodynamic simulations of \citet{OppenheimerDave2010} suggest that the recycling of gas blown out by winds is the dominant mode of accretion in halos with masses above $10^{11.2} M_{\odot}$ at $z\sim0$, with recycling times $<1$ Gyr.  Further, detailed simulations of individual galaxy halos show that accretion occurs in the plane of the galactic disk, rather than along the minor axis (\citealt{Stewart2011,Shen2011}, M\'enard \& Murray 2011, in prep).  Winds vent out of the galaxy along the path with the lowest ambient gas pressure, i.e., the minor axis, preventing accretion from occurring in locations other than along the disk plane.  While we have not ruled out the presence of infall onto galaxies along their minor axes, results from these simulations are fully consistent with our detection of inflows along the line-of-sight toward several highly-inclined, disk-dominated galaxies.

These results highlight the importance of analysis of cool gas kinematics in distant galaxies on an individual basis.  
Only through examination of absorption line profiles for a large sample of objects were a handful of examples of cool gas accretion, a process fundamental to galaxy formation, identified.
At the same time, these spectra provide only a cursory view of the complexities of gas infall and recycling.  Studies of gas kinematics at higher spectral resolution \citep[e.g.,][]{Pettini2001,Dessauges-Zavadsky2010} and in large galaxy samples are needed to achieve tighter constraints on the frequency of inflows and the concurrent action of outflows and inflows. Equally propitious are studies of cool gas abundances along sightlines to background QSOs that may differentiate between pristine gas accreted from the IGM and material that has been previously recycled \citep{Ribaudo2011}.  The combination of these experiments in studies of individual halos will in turn enable the simultaneous mapping of gas abundances and kinematics relative to the host galaxies.  
In concert with hydrodynamic simulations that track the accretion, expulsion, and recycling of gas, these observations will provide unprecedented insight into the processes regulating galaxy growth.

\acknowledgements
The authors are grateful for support for this project from NSF grants
AST-0808133, AST-0507483, and AST-0548180.
We thank Robert da Silva, Crystal Martin, Joop Schaye, Greg Stinson, Ben Weiner, Arjen van der Wel and Luke Winstrom for helpful discussions 
of this analysis.



\begin{deluxetable}{llcccccc}
\tabletypesize{\footnotesize}
\tablecolumns{7}
\tablecaption{Galaxy Properties and Inflow Measurements\label{tab.inflows}}
\tablewidth{0pt}
\tablehead{\colhead{} & \colhead{} & \colhead{TKRS3974} & \colhead{TKRS4045} & \colhead{TKRS4387} & \colhead{EGS12027936} & \colhead{J033249.5} & \colhead{J033227.8}} \\
\startdata
\underline{Galaxy Properties} & & & & & & & \\
Right Ascension & & 12:37:01.65 & 12:36:39.70 & 12:36:54.99 & 14:19:26.49 & 03:32:49.52 & 03:32:27.83 \\
Declination & & +62:18:14.3 & +62:15:26.1 & +62:16:58.2 & +52:46:09.4 & -27:46:29.9 & -27:55:48.8 \\
z \tablenotemark{a} & &    0.43708 &    0.37687 &    0.50334 &    1.03847 &    0.52313 &    0.66424 \\
$M_B$ (mag) & & -20.90 & -20.61 & -21.36 & -22.74 & -21.19 & -21.50 \\
$U - B$ (mag) & &   0.91 &   1.00 &   0.93 &   0.50 &   0.60 &   0.65 \\
$R$ (kpc) & &    3.2 &    5.1 &    4.7 &    3.9 &    2.6 &    5.2 \\
$R$ (arcsec) & &    0.6 &    1.0 &    0.8 &    0.5 &    0.4 &    0.7 \\
$i$ (deg) & &   56.8 &   67.3 &   57.2 &   56.6 &   38.1 &   67.2 \\
$\mathrm{SFR}$ ($M_{\odot}~\mathrm{yr}^{-1}$) & & $   1.8^{+   0.2}_{-   0.2} $ & $   1.3^{+   0.2}_{-   0.2} $ & $   1.5^{+   0.3}_{-   0.2} $ & $  41.4^{+  28.1}_{-  14.0} $ & $  11.3^{+   2.9}_{-   5.8} $ & $  18.1^{+   0.2}_{-   3.6} $ \\
$\log M_*/M_{\odot}$ & & $ 10.26^{+  0.09}_{-  0.06} $ & $ 10.30^{+  0.05}_{-  0.06} $ & $ 10.47^{+  0.08}_{-  0.09} $ & $ 10.14^{+  0.06}_{-  0.04} $ & $  9.59^{+  0.25}_{-  0.04} $ & $  9.83^{+  0.01}_{-  0.03} $ \\
\tableline \\ [-1.5ex]
\underline{Transition} & \underline{Inflow Measurements}\tablenotemark{b} & & & & & & \\
\ion{Mg}{2} 2796, 2803 & $P_\mathrm{in,1}$ & $1.000$ & $0.959$ & $0.995$ & \nodata & $0.997$ & $1.000$ \\
 & $\Delta v_\mathrm{1}$ ($\mathrm{km~s^{-1}}$) & $   164^{+    20}_{   -20} $ & $   160^{+    81}_{   -91} $ & $    71^{+    22}_{   -24} $ & \nodata & $    54^{+    21}_{   -19} $ & $    97^{+    19}_{   -19} $ \\
 & $\rm EW_\mathrm{1}$ ($\mathrm{\AA}$) & $   3.2^{+   0.4}_{  -0.4} $ & $   3.3^{+   0.9}_{  -0.8} $ & $   3.1^{+   0.4}_{  -0.4} $ & \nodata & $   1.7^{+   0.3}_{  -0.2} $ & $   2.9^{+   0.4}_{  -0.4} $ \\
[-1.5ex] \\ \cline{2-8} \\ [-1.5ex]
 & $P_\mathrm{in,2}$ & $0.986$ & $0.766$ & $0.970$ & \nodata & $0.951$ & $1.000$ \\
 & $\Delta v_\mathrm{2}$ ($\mathrm{km~s^{-1}}$) & $   201^{+    34}_{   -40} $ & N/C & $   125^{+    60}_{   -54} $ & \nodata & $   193^{+   109}_{   -93} $ & $   177^{+    46}_{   -46} $ \\
 & $\rm EW_\mathrm{2}$ ($\mathrm{\AA}$) & $   2.5^{+   0.6}_{  -0.5} $ & $   0.7\pm   0.3$ & $   2.4^{+   1.9}_{  -0.9} $ & \nodata & $   0.6^{+   0.4}_{  -0.4} $ & $   1.8^{+   0.5}_{  -0.5} $ \\
 & $\log N_\mathrm{MgII}$ ($\mathrm{cm}^{-2}$) & $ >   14.3$ & N/C & $ >   13.9$ & \nodata & $ >   11.0$ & $ >   13.8$ \\
 & $dM_\mathrm{in}/dt$ ($M_{\odot}~\mathrm{yr}^{-1}$) &      0.69 & N/C &      0.24 & \nodata &   0.00030 &      0.29 \\
\tableline \\ [-1.5ex]
\ion{Fe}{2} 2586, 2600 & $P_\mathrm{in,1}$ & \nodata & \nodata & \nodata & $1.000$ & $1.000$ & $1.000$ \\
 & $\Delta v_\mathrm{1}$ ($\mathrm{km~s^{-1}}$) & \nodata & \nodata & \nodata & $    70^{+    13}_{   -13} $ & $    82^{+    23}_{   -22} $ & $    87^{+    16}_{   -15} $ \\
 & $\rm EW_\mathrm{1}$ ($\mathrm{\AA}$) & \nodata & \nodata & \nodata & $   3.1^{+   0.2}_{  -0.2} $ & $   1.2^{+   0.2}_{  -0.2} $ & $   2.4^{+   0.3}_{  -0.3} $ \\
[-1.5ex] \\ \cline{2-8} \\ [-1.5ex]
 & $P_\mathrm{in,2}$ & \nodata & \nodata & \nodata & $0.983$ & $0.660$ & $0.990$ \\
 & $\Delta v_\mathrm{2}$ ($\mathrm{km~s^{-1}}$) & \nodata & \nodata & \nodata & $   192^{+    41}_{  -118} $ & N/C & $   128^{+    33}_{   -32} $ \\
 & $\rm EW_\mathrm{2}$ ($\mathrm{\AA}$) & \nodata & \nodata & \nodata & $   1.4^{+   1.3}_{  -0.4} $ & $   0.6\pm   0.1$ & $   1.3^{+   0.7}_{  -1.3} $ \\
 & $\log N_\mathrm{FeII}$ ($\mathrm{cm}^{-2}$) & \nodata & \nodata & \nodata & $ >   14.6$ & N/C & $ >   14.9$ \\
 & $dM_\mathrm{in}/dt$ ($M_{\odot}~\mathrm{yr}^{-1}$) & \nodata & \nodata & \nodata &       1.9 & N/C &       2.8 \\
\enddata
\tablenotetext{a}{Estimated statistical and systematic uncertainty is $28\mkms$.  The redshift of EGS12027936 was determined by the DEEP2 Survey, which reports a $1\sigma$ redshift uncertainty of $36\mkms$ \citep{Weiner2009}.}
\tablenotetext{b}{``N/C" (for ``not constrained") is used in cases for which less than 95\% of the PPDF for the velocity offset lies at $> 0\mkms$ (i.e., $P_\mathrm{in} < 0.95$).  Measurements subscripted with ``1" refer to results from one-component model fitting (\S\ref{sec.specanalysis}).  Measurements subscripted with ``2" report two-component model results, except in cases with $P_\mathrm{in,2} < 0.95$, where they list the EW of absorption measured in the velocity range $30 \mkms < v < 300 \mkms$.  Errors report the $\pm 34\%$ probability intervals on each quantity.}
\tablecomments{Galaxies are named according to ID numbers listed in \citet[][TKRS]{Wirth2004}, the AEGIS survey \citep[][EGS]{Davis2007}, and \citet[][J0332...]{Giavalisco2004}. SFRs and $M_*$s are calculated by fitting model spectral energy distributions to the broad-band photometry listed in \S\ref{sec.observations} along with the photometry of \citet{Bundy2006} using the code MAGPHYS \citep{daCunha2008}.}
\end{deluxetable}

\end{document}